\documentclass[journal]{IEEEtran}

\usepackage{xfrac}
\usepackage{amssymb}
\usepackage{graphicx}
\usepackage{amsmath}
\usepackage{subfigure}
\usepackage{url}
\usepackage{epstopdf}
\usepackage{cite}
\usepackage{verbatim}
\usepackage{amstext}
\usepackage{enumerate}
\usepackage{algorithmic}
\usepackage{amsfonts}
\usepackage{algorithm}
\usepackage{float}
\usepackage{soul,color}
\DeclareMathOperator*{\argmin}{\arg\!\min}
\DeclareMathOperator*{\argmax}{\arg\!\max}
\newcommand{\V}[1]{\mathbf{#1}}
\newcommand {\vc} [1] {\boldsymbol{#1}}

\begin{document}

\title{Constrained State Estimation - A Review}
%\author{Ghulam~Rasool (rasool@rowan.edu)}

\author{Nesrine Amor, Ghualm Rasool, and Nidhal C. Bouaynaya 
       % <-this % stops a space
\thanks{N. Amor is with the National Superior School of Engineers of Tunis (ENSIT), University of Tunis, Tunis, Tunisia, e-mail: (nisrine.amor@hotmail.fr}
\thanks{G. Rasool and N. C. Bouaynaya are with Department of Electrical and Computer Engineering, Rowan University, New Jersey, USA, e-mail: (rasool@rowan.edu, bouaynaya@rowan.edu)}}% <-this %
%,~\IEEEmembership{Student Member,~IEEE,}
%Nidhal~Bouaynaya,~\IEEEmembership{Member,~IEEE} and~Roman~Shterenberg

%\thanks{Ghulam Rasool is with the Department of Electrical and Computer Engineering of the Rowan University email: rasool@rowan.edu.}
\maketitle
%\thanks{N. Bouaynaya is an Assistant Professor with the Electrical and Computer Engineering Department, Rowan University, Glassboro, NJ-08028, USA (email: bouaynaya@rowan.edu).}
%\thanks{Information about Dr. Roman.}
%}

\begin{abstract}
The real-world applications in signal processing generally involve estimating the system state or parameters in nonlinear, non-Gaussian dynamic systems. The estimation problem may get even more challenging when there are physical constraints on the system state. This tutorial-style paper reviews the Bayesian state estimation for (non)linear state-space systems and introduces the formulation of constrained state estimation in such scenarios. Specifically, we start by providing a review of unconstrained state estimation using Kalman filters (KF) for the linear systems and their extensions for nonlinear state-space systems, including extended Kalman filters (EKF), unscented Kalman filters (UKF), and ensemble Kalman filters (EnKF). Next, we present particle filters (PFs) for nonlinear state-space systems. Finally, we review constrained state estimation using various filtering techniques and highlight the advantages and disadvantages of the different constrained state estimation approaches.
\end{abstract}
%\begin{IEEEkeywords}

%\end{IEEEkeywords}
%\IEEEpeerreviewmaketitle

\section{Introduction}
% Motivation
%The state of many dynamical systems is often required to satisfy certain constraints arising from basic physical laws, mathematical properties or geometric considerations of the underlying system, e.g., maximum power or transmission capacity, energy conservation laws and bounded parameters. In fact, constrained systems are already omnipresent in many real-world applications including camera tracking \cite{Julier_TwoStep_2007}, fault diagnosis \cite{Simon_2005, Simon_2006, Simon_PDFTrunc_2010}, chemical processes \cite{Vachhani_RNDDR_2005, Vachhani_RNDDR_2006}, vision-based systems \cite{Porrill_PerfMeas_1988}, target tracking \cite{Wang_PerfMeas_2002, Xu_2013}, biomedical systems \cite{Chia_1991, Rasool_2013}, robotics \cite{Spong_2006}, navigation \cite{Simon_Projection_2002}, tracking of ground vehicles \cite{Yang_2005}, and tracking and detection in a maritime scenario \cite{Battistello_2011}. Using explicit constraints, rather than their implicit inclusion through penalty and barrier methods, simplifies the design specification to focus on the performance objective.
The states of many dynamical systems are confined within constrained regions owing to relevant physical laws, geometric considerations, and kinematic limits, such as material balance, maximum power rating, bounds on the output of actuators and plants, and speed constraints in road networks \cite{amor2019constrained, albers2019ensemble}. The mathematical formulation of such constraints involves defining a set of (non)linear relationships in the form of (in)equalities constraining the state estimates. One possible solution is to incorporate the constraints into the state-space model of the system. However, generally, that is not possible without a significant increase in the complexity of the model. During the state estimation process, taking constraints into account leads to accurate and physically relevant estimates. For instance, the exploitation of a known road network has been proven effective in the tracking of ground vehicles \cite{yang02}. Similarly, in a maritime scenario, the knowledge of shipping lanes and sea/land distinction can improve the tracking and detection performance \cite{batt02}.

The state-space representation provides an extremely flexible framework for modeling discrete-time dynamical systems using two mathematical relationships, i.e., the state transition and observation. The former captures the evolution of the hidden state over time, and the latter provides the noisy measurement of a (non)linear function of the state. The descriptive power of the state-space representation comes at the expense of intractability. Generally, it is not possible to obtain an analytic solution to the state estimation problem except for a small number of cases, e.g., linear systems with Gaussian noise.

%In the Bayesian framework, inference of the hidden state given all available observations at that time relies upon the posterior density function (pdf). For the linear and Gaussian estimation problems, optimal solution can be obtained by the Kalman filter. For nonlinear and non-Gaussian state-space models, Optimal estimation problems do not typically admit analytic solutions to the posterior density estimation problem. Therefore, a numerical method is needed to approximate the posterior pdf of the state vector.
In the Bayesian framework, the inference of the hidden state given all available observations relies upon estimating the posterior density function (pdf). For the linear systems with Gaussian noise, the closed-form optimal solution is given by the Kalman filter (KF). On the other hand, for nonlinear and non-Gaussian state-space models, two fundamental techniques have been emerging, parametric and nonparametric \cite{thrun2002probabilistic}. The parametric methods include the extended Kalman filter (EKF), unscented Kalman filter (UKF), ensemble Kalman filter (EnKF), and moving horizon estimation. The nonparametric techniques are based on sequential Monte Carlo methods and include particle filters (PF) \cite{thrun2002probabilistic}.

%Bayesian estimation is a framework for the formulation of statistical inference problems. Dynamic systems are modeled using state evaluation and observation relations. The former captures the evolution of the state with time and later provides noisy measurement of a probably nonlinear function of the state. However, some additional information about the system may be available due to the underlying physical process. During state estimation, such information can be used to improve the state estimate. Such schemes fall under the umbrella of ``Constrained Bayesian Estimation''.

This paper provides a tutorial-style review of constrained state estimation methods for KF, EKF, UKF, EnKF, moving horizon estimation, and PF. In Section \ref{section-problem}, we start by presenting the problem statement. Section \ref{section-uncontrained-PF} provides detailed description of unconstrained Bayesian state estimation techniques. Section \ref{section-constrained-CBE} presents the literature available in constrained state estimation for KF, EKF, EnKF, and moving horizon estimation. In Section \ref{section-constrained-PF}, we formulate and review of the constrained PF problem. Finally, Section \ref{section-conclusion} summarizes and concludes the paper.

\section{Problem Statement} \label{section-problem}
\subsection{System Definition}
We consider a general state-space representation defined by the state transition and measurement models for discrete-time systems, given by:
\begin{align}
\vc{x}_{t+1} = \V{f}_t(\vc{x}_t) + \vc{w}_t, \label{eqnSystemDynamics} \\
\vc{y}_{t} = \V{h}_t (\vc{x}_t) + \vc{v}_t, \label{eqnMeasurement}
\end{align}
where $\vc{x}_{t}\in \mathbb{R} ^{n_x}$ and $\vc{y}_{t}\in \mathbb{R} ^{n_y}$ represent the hidden state and observation at time $t$, respectively, $ t \in \mathbb{N} $ represents the time index and $ n_x $ and $ n_y $ are state and output dimensions. In out settings, $ \V{f}_t : \mathbb{R}^{n_x} \to \mathbb{R}^{n_x} $ and $ \V{h}_t : \mathbb{R}^{n_x} \to \mathbb{R}^{n_y} $ are (non)linear mappings, , and $ \vc{w}_t $ and $ \vc{v}_t $ are zero-mean process and measurement noise with known PDFs $ p(w) $ and $ p(v) $. Both noise sequences are uncorrelated with each other and the initial condition of the state $ x_0 $ given by $ p(x_0) $.

The mapping functions $ \V{f}_t(\vc{x}_t) $ and $ \V{h}_t(\vc{x}_t) $ can be defined in term of PDFs, i.e.,
\begin{align}
p(\vc{x}_t|\vc{x}_{t-1}) &= p_{w_t}(\vc{x}_t - \V{f}_t(\vc{x}_{t-1})), \\
p(\vc{y}_t|\vc{x}_{t}) &= p_{v_t}(\vc{y}_t - \V{h}_t(\vc{x}_t)).
\end{align}

%\subsection{Bayesian Recursion}
\subsection{Optimal State Estimation}
The state estimation problem aims at finding the hidden state $ \vc{x}_t $ using all available measurements up to time $ t $, $\vc{y}_{1:t}=[\vc{y}_{1},...,\vc{y}_{t}]$. The solution to this problem is the density of the system state conditioned on the measurements; either joint PDF $ p(\vc{x}_1, \ldots, \vc{x}_t | \vc{y}_{1:t}) $ or the marginal PDF  $ p(\vc{x}_t | \vc{y}_{1:t})$ \cite{thrun2002probabilistic}.

We make two assumptions here, (1) the system state follows a first-order Markov process, i.e., $ p(\vc{x}_t | \vc{x}_{1:t}, \vc{y}_{1:t}) = p(\vc{x}_t|\vc{x}_{t-1}) $ and 2) measurements are conditionally independent given the system state, i.e., $ p(\vc{y}_t| \vc{x}_{1},...,\vc{x}_{t}, \vc{y}_{1},...,\vc{y}_{t-1}) = p(\vc{y}_t|\vc{x}_t) $.

In Bayesian estimation framework, a recursion can be defined to estimate posterior PDF $  p(\vc{x}_t|\vc{y}_{1:t}) $ using the prior PDF $  p(\vc{x}_{t-1}| \vc{y}_{1:t-1}) $, transition PDF $ p(\vc{x}_t|\vc{x}_{t-1}) $ and the likelihood PDF $ p(\vc{y}_t|\vc{x}_t) $. We consider the marginal posterior, however, the same results hold true for the joint conditional PDF also. Using Bayes rule and Chapman-Kolmogorov equation, the posterior PDF can be computed recursively using the following two-step relationship \cite{thrun2002probabilistic}:

Prediction Step
\begin{align} 
&p(\vc{x}_t|\vc{y}_{1:t-1}) =\int_{}^{} p(\vc{x}_{t-1}|\vc{y}_{1:t-1})  ~p(\vc{x}_{t}|\vc{x}_{t-1}) ~d\vc{x}_{t-1},  \label{eq3a} 
\end{align}

Update Step
\begin{align}
    p(\vc{x}_t|\vc{y}_{1:t}) & = \frac{p(\vc{y}_{1:t}|\vc{x}_t)p(\vc{x}_t)}{p(\vc{y}_{1:t})}, \nonumber\\
    & = \frac{p(\vc{y}_t|\vc{y}_{1:t-1}, \vc{x}_t) p(\vc{y}_{1:t-1}|\vc{x}_t) p(\vc{x}_t)} {p(\vc{y}_t|\vc{y}_{1:t-1}) p(\vc{y}_{1:t-1})}, \nonumber \\
    & = \frac{p(\vc{y}_t|\vc{x}_t) p(\vc{x}_t|\vc{y}_{1:t-1})}{p(\vc{y}_t|\vc{y}_{1:t-1})}, \nonumber \\
    & = \frac{p(\vc{y}_t|\vc{x}_t) p(\vc{x}_t|\vc{y}_{1:t-1})}{\int p(\vc{y}_t | \vc{x}_t) p(\vc{x}_t | \vc{y}_{1:t-1}) d\vc{x}_t}. \label{eq3b}
\end{align}

%\begin{align}
%&p(\vc{x}_t|\vc{y}_{1:t}) = \frac{p(\vc{y}_t |\vc{x}_t) ~p(\vc{x}_{t}|\vc{y}_{1:t-1})} {\int_{}^{} p(\vc{y}_t |\vc{x}_t) ~p(\vc{x}_{t}|\vc{y}_{1:t-1}) ~d\vc{x}_{t}}. \label{eq3b}
%\end{align}
For a general case with nonlinear state transition $ \V{f}_t(\vc{x}_t) $ and observation functions $ \V{h}_t(\vc{x}_t) $ and non-Gaussian noise sequences $\vc{w}_t$ and $\vc{v}_t$, eqs. (\ref{eq3a}) and (\ref{eq3b}) are only a conceptual solution owing to the intractable integrals. For the case of linear systems with Gaussian noise, a closed-form optimal solution in the form of KF exists \cite{Simo06}.

\section{Unconstrained State Estimation} \label{section-uncontrained-PF}
\subsection{Kalman Filters for Linear Systems with Additive Gaussian Noise}
%Generally, for tracking problems with linear and Gaussian models, an optimal solution can be obtained using the Kalman Filter. In fact, many real-world applications such as target tracking, electric power systems, navigation and biomedical engineering, these linear and Gaussian assumptions do not hold. Furthermore, approximations are required for these nonlinear and non-Gaussian tracking problems. Thus, many approaches were introduced to solve this problem such as Extended Kalman Filter (EKF), Moving Horizon estimation (MHE), Ensemble Kalman Filter (EnKF), Unscented Kalman Filter (UKF) and Particle Filters (PF). In this section, we introduce these filters for linear and nonlinear systems when there are no constraints on the system.
%In this section, we introduce the Kalman filter for linear system and extended Kalman filter, unscented Kalman filter, moving horizon estimation and ensemble Kalman filter for nonlinear system when there are no constraints on the system.

The state transition function $ \V{f}_t(\vc{x}_t) $ and the observation $ \V{h}_t(\vc{x}_t) $ function are both linear functions and are represented by matrices $F_t \in \mathbb{R} ^{n_x \times n_x} $ and $H_t \in \mathbb{R} ^{n_y \times n_x}$ respectively. The noise sequences are both additive and Gaussian, and defined as $ w_t \sim \mathcal{N}(\V{0},Q_t)$ and $ v_t \sim \mathcal{N}(\V{0},R_t)$. The initial system state $\vc{x}_0$ is also assumed to be known with a Gaussian distribution $\mathcal{N}(x_0,P_0)$. The Kalman filter, a minimum mean square error (MMSE) estimator, is defined with two steps, the prediction step and the update step. We use $\hat{}$ to represent the estimate, $^T$ for the matrix transpose operation, and $^-$ and $^+$ symbols present quantities before and after a state or observation update is applied.

\begin{subequations} \label{eqnKFPrediction}
\begin{align}
\intertext{Prediction step}
        \hat{\vc{x}}^-_{t} & = F_{t-1} \hat{x}^+_{t-1}, \label{eqnKFPrediction:a}\\
        P^-_t & = F_{t-1} P^+_{t-1}F^T_{t-1}+Q_t, \label{eqnKFPrediction:b}\\
        P_{xy} & = P^-_t H^T_t, \label{eqnKFPrediction:c}\\
        P_y & = H_t P^-_tH^T_t+R_t, \label{eqnKFPrediction:d} 
\end{align}
\end{subequations}
\begin{subequations} \label{eqnKFUpdata}
\begin{align}
\intertext{Update step}
          \hat{y}_t & = H_t\hat{x}^-_t, \label{eqnKFUpdata:a} \\
          K_t & = P_{xy}P_y^{-1}, \label{eqnKFUpdata:b} \\
 \hat{x}_t^+  & = \hat{x}^-_t + K_t(y_t - \hat{y}_t), \label{eqnKFUpdata:c} \\
        P_t^+ & = P^-_t - P_{xy}P_y^{-1}P_{xy}^T, \label{eqnKFUpdata:d}
\end{align}
\end{subequations}
where $K$ represents the Kalman gain matrix. We can show that KF is the best linear unbiased estimator (BLUE) for all linear systems with additive noise.

\subsection{Extended Kalman Filter (EKF)}
The EKF assumes that the nonlinear state transition $\V{f}_t(\vc{x}_t)$ and/or observation $\V{h}_t(\vc{x}_t)$ functions can be linearized using the Taylor series \cite{Jazwinski_Book_1970}, \cite{Simon_Book_2006}.
\begin{align} \label{eq:linearizzation}
F_t = \frac{\partial \V{f}_t(\vc{x}_t)}{ \partial \vc{x}_t } \biggr|_{\hat{\vc{x}}^+_t} \text{ and } H_t = \frac{\partial \V{h}_t(\vc{x}_t)}{ \partial \vc{x}_t } \biggr|_{\hat{\vc{x}}^-_t}. \end{align}

The error covariance $P_t$ is propagated (using eqs. \ref{eqnKFPrediction:b}, \ref{eqnKFPrediction:c}, and \ref{eqnKFPrediction:d}) with the linearized functions $F_t$ and $H_t$ from eq. (\ref{eq:linearizzation}). The state estimate $\hat{\vc{x}}^-_{t}$ (eq. \ref{eqnKFPrediction:a}) and $\hat{y}_t$ (eq. \ref{eqnKFUpdata:a} ) are calculated using nonlinear functions $\V{f}_t(\vc{x}_t)$ and $\V{h}_t(\vc{x}_t)$ as given below:

\begin{align}
    \hat{\vc{x}}^-_{t} & = \V{f}_t (\hat{x}^+_{t-1}), \\
    \hat{\vc{y}}_t & = \V{h}_t(\hat{\vc{x}}^-_t).
\end{align}

The computational complexity for the calculation of $F_t$ and $H_t$ matrices may prohibit the use of EKF in some applications \cite{Simon_Book_2006}. Furthermore, the linearization of nonlinear systems may introduce errors in the estimation of the state, and in the worst-case, the filter may diverge especially for highly nonlinear systems \cite{Julier_UKF_2004}.

\subsection{Unscented Kalman Filter (UKF)}
UKF addresses the issues raising from the approximation operations and limit the performance of EKF \cite{wan2000unscented}. UKF uses a set of carefully chosen deterministic samples, called \emph{sigma points}, to propagate the mean and covariance of the posterior distribution of the state. The set of sigma points capture the true mean and covariance of the posterior for the case of Gaussian distribution when propagated through the nonlinear system. 

Considering the system state $\vc{x}_t$ with mean $\hat{\vc{x}}_t$ and covariance $P_t$, we can chose sigma points $ \mathcal{X}_{j,t} \in \mathbb{R}^{n_x} $, $ j=0,1,\ldots,2n $ using following relationships \cite{Julier_UKF_1997, Julier_UKF_2004, Julier_UKF_2000, wan2000unscented}:

\begin{align} 
\mathcal{X}_{0,t} &= \hat{\vc{x}}_t, \label{eqnUTPoints:1} \\
\mathcal{X}_{j,t} &= \hat{\vc{x}}_t + \left( \sqrt{(n + \lambda) P_t} \right)_j  ~~~    j = 1, \ldots, n \\
\mathcal{X}_{j,t} &= \hat{\vc{x}}_t - \left( \sqrt{(n + \lambda) P_t} \right)_{i-n}      j = n+1, \ldots, 2n \\
W_0^{(m)} &= \frac{\lambda}{n+\lambda}, \\
W_0^{(c)} &= \frac{\lambda}{n+\lambda} + (1-\alpha^2 +\beta), \\
W_j^{(m)} &=W_j^{(c)} = \frac{1}{2(n+\lambda)}, ~~~ j=1,\ldots,2n,
\end{align}
where $\left( \sqrt{(n + \lambda) P_t} \right)_i$ represents the $j$th row of the matrix square root. $\lambda = \alpha^2 (n+\kappa) - n$ is a scaling parameter. $\alpha$ determines the spread of the sigma points around the mean $\hat{\vc{x}}_t$ and is set to a small value, e.g., $1e-3$ \cite{wan2000unscented}. $\kappa$ is secondary scaling parameter and is generally set to $0$. $\beta$ encodes the prior knowledge about the distribution of $\vc{x}$ and is set to $2$ for the Gaussian distribution \cite{wan2000unscented}.  

The sigma points $ \left[\mathcal{X}_{j,t}\right]_{j=0}^{2n} $ (n-dimensional vectors) are passed through the nonlinear function, i.e., 
\begin{align}
    \mathcal{Y}_{j,t} = \V{g}(\mathcal{X}_{j,t}), ~~~ j=1, \ldots, 2n,  
\end{align}
where $\V{g}$ is a nonlinear function, e.g., state transition $\V{f}$ or observation $\V{h}$. The mean and covariance after the transformation can be calculated using:
\begin{align}
    \hat{\vc{y}} &\approx \sum_{j=0}^{2n} W_j^{(m)} \mathcal{Y}_j, \\
    P_y &\approx \sum_{j=0}^{2n} W_j^{(c)} \left[ \mathcal{Y}_j - \hat{\vc{y}} \right] \left[ \mathcal{Y}_j - \hat{\vc{y}} \right]^T. \label{eqnUTPoints:cov}
\end{align}
The mathematical relations described in eqs. (\ref{eqnUTPoints:1}) - (\ref{eqnUTPoints:cov}) are referred to as unscented transformation (UT).

UKF uses UT (eqs. \ref{eqnUTPoints:1} - \ref{eqnUTPoints:cov}) and KF relations (eqs. \ref{eqnKFPrediction} and \ref{eqnKFUpdata}) to estimate the state mean and covariance without having to linearize the nonlinear functions $\V{f}_t and \V{h}_t$ and calculating the Jacobian matrices. However, UKF has its own limitations, e.g., it may not work well with systems that have nearly singular covariance matrices. This problem is linked to the matrix square root operation performed on the covariance matrices using Cholesky decomposition, which, in turn, can be computationally demanding also. It is also important to mention that all three filters described above, i.e., KF, EKF and UKF assume that the \emph{prior} distribution function follows Gaussian distribution.
The approximations obtained with at least $ 2n+1 $ sampling points are accurate to the 3\textsuperscript{rd} order for the additive Gaussian noise and  for all types of nonlinear functions and at least to the 2\textsuperscript{nd} for non-Gaussian inputs \cite{Terejanu_UKF_Tutorial_2009}. 

\subsection{Moving Horizon Estimation (MHE)}
Moving horizon estimation (MHE) is an optimization approach that can be used to estimate the unknown system state \cite{Rao_MHE_2001}, \cite{Rao_MHE_2002}, \cite{Rao_MHE_2003}, \cite{Shao_JProcCon_Optimization_2010}. MHE employs an iterative procedure that relies on linear or nonlinear programming to find the desired solution, i.e., the system state. For maximum \emph{a posteriori} estimate, the state estimation problem can be expressed as:
\begin{equation} \label{eqnMAPBayesian}
\hat{x}_t :=\argmax\limits_{x_t} p(x_t|Y_t).
\end{equation}
For a moving horizon of size $ h \in \{0,t\} $, we can determine
\begin{align}
{\hat{x}_{t-h}, \ldots,\hat{x}_{t}} & = \argmax\limits_{x_{t-h}, \ldots,x_{t}} p(x_{t-h}, \ldots,x_{t} | Y_t).
\end{align}
Using Markovian assumption and Bayes rules, we get
\begin{align} \label{eqnMHE1}
& p(x_{t-h}, \ldots,x_{t} | Y_t) \propto \nonumber\\
& \prod_{j=t-h}^{t}p(y_j|x_j) \prod_{j=t-h}^{t-1} p(x_{j+1}|x_j) p(x_{t-h}|Y_{t-h-1}).
\end{align}
With Gaussian assumption and taking logarithm, we have
\begin{align} \label{eqnMHE}
\argmin\limits_{x_{t-h},\ldots,x_t}& \sum_{j=t-h}^{t-1}\|y_j - \V{h}_j(x_j)\|_{R_j^{-1}}^2 + \nonumber\\
& \|x_{j+1} - \V{f}_j(x_j)\|_{Q_j^{-1}}^2 + \|x_{t-h} - \bar{x}_{t-h}\|_{\Pi_{t-h}^{-1}}^2,
\end{align}
where the last term is the arrival cost and for $ t=h, \;\; \Pi_{t-h} = \Pi_0 $, i.e., the initial covariance of the state estimate at time $ t=0 $. 

One motivation for the development of MHE was to formulate a mathematical optimization problem where constraints on the system state can be naturally incorporated into the estimation (now optimization) framework \cite{Patwardhan_Review_2012}.

\subsection{Ensemble Kalman Filter (EnKF)}
The ensemble Kalman filter (EnKF) belong to the a broader class of sequential Monte Carlo methods \cite{Gillijns_EnKF_2006}, \cite{Evensen_EnKF_2009}, \cite{Burgers_EnKF_1998}. In the most general form, an EnKF is based on the premise that it is sufficient to estimate first two moments of the probability density functions of interest, i.e., $ p(x_t|Y_{t-1}) $ and $ p(x_t|Y_{t}) $, for the time-update and measurement-update steps \cite{Prakash_C_EnKF_2010}.

EnKF is initialized with $ N $ samples (or particles) that are sampled from a given probability distribution function. At each subsequent time step, $ N $ samples are drawn from process and observation noise distribution functions and propagated through system dynamics to compute a set of transformed particles. 

Given that we have $ N $ particles $ \{\hat{x}^{i,+}_{t-1} \}_{i=1}^{N} $,
\begin{subequations}
\begin{align}
\intertext{Prediction step}
  \hat{x}^{i,-}_{t} & = \V{f}_{t-1} \hat{x}^{i,+}_{t-1} + w_t^i \\
  \hat{y}^{i,-}_{t} & = \V{h}_{t}(\hat{x}^{i,-}_{t}) + v_t^i \\
        \bar{x}^-_t & =\frac{1}{N} \sum\limits_{i=1}^{N} \hat{x}^{i,-}_{t} \\
        \bar{y}^-_t & = \frac{1}{N} \sum\limits_{i=1}^{N} \hat{y}^{i,-}_{t} \\
             P_{xy} & = \frac{1}{N-1} \sum\limits_{i=1}^{N}[\hat{x}^{i,-}_{t} - \bar{x}^-_t] [\hat{y}^{i,-}_{t} - \bar{y}^-_t]^T \\
             P_{y} & = \frac{1}{N-1} \sum\limits_{i=1}^{N}[\hat{y}^{i,-}_{t} - \bar{y}^-_t] [\hat{y}^{i,-}_{t} - \bar{y}^-_t]^T
\end{align}
\end{subequations}
\begin{subequations}
\begin{align}
\intertext{Update step}
          K_t & = P_{xy}P_y^{-1}  \\
 \hat{x}_t^{i,+}  & = \hat{x}^{i,-}_t + K_t[y_t - (\V{h}_{t}(\hat{x}^{i,-}_{t}) + v_t^i)] \\
 \bar{x}_t^{i,+} & = \frac{1}{N} \sum\limits_{i=1}^{N} \hat{x}_t^{i,+} \\
 P_t & = \frac{1}{N-1} \sum\limits_{i=1}^{N} [\hat{x}_t^{i,+} - \bar{x}_t^{i,+}][\hat{x}_t^{i,+} - \bar{x}_t^{i,+}]^T
\end{align}
\end{subequations}

The estimate accuracy of the EnKF depends on the number of samples $N$ \cite{Gland_EnKF_2009}. The difference between EnKF and particle filters is that EnKF makes the assumption that all probability density functions are Gaussian and can be represented only by the mean and covariance. EnKF can be used for systems with non-Gaussian probability distribution functions. However, as only first two moments are being used in the prediction and the update steps, the estimation results may not be accurate for systems that do not follow the Gaussian assumption.

Various methods, including EKF, UKF, MHE, and EnKF may not accurately estimate systems state when the underlying system dynamics are highly nonlinear and/or the noise in the system does not follow Gaussian distribution. particle filters (PFs) are able to handle nonlinear systems with non-Gaussian noise. PFs are flexible and simple simulation-based numerical approaches used for estimating the system state in a sequential manner.

\subsection{Particle Filters (PFs)}
PFs solve the optimal estimation problem in nonlinear and non-Gaussian dynamic systems by incorporating sequential Monte Carlo sampling within the Bayesian filtering framework \cite{Doucet_2000} \cite{Doucet_2009} \cite{Gordon_1993}, \cite{Doucet_Book_2001} \cite{Merwe_Thesis_2004}. The goal is to estimate the posterior density for the state using Bayesian recursion.
PFs approximate the posterior PDF using a group of samples (also called particles) $ x_t^i $ and their associated weights $ w_t^i \ge 0 $ as:

\begin{equation}  \label{eq5a}
\hat{p}(x_t|Y_t) = \sum_{i=1}^{N} w_t^{(i)} \delta(x_t - x_t^{(i)}),
\end{equation}
where $ \delta(.) $ represents the Dirac delta function and $ N $ is the number of particles.

The conditional mean estimate of the state is given by the weighted mean of the particles as follows:
\begin{equation} \label{eq7}
\widehat{\vc{x}}_t = {E[\vc{x}_t|\vc{Y}_{t}]} \approx\sum\limits_{i=1}^N w^{(i)}_t \vc{x}_t^{(i)}.
\end{equation}

The particles are required to be sampled from the true posterior PDF, which is not available. Therefore, another PDF, usually referred to as the \emph{importance distribution} or the \emph{proposal distribution} $q(\vc{x}_t|\vc{x}_{t-1})$, which is generally easy to sample from, is defined \cite{Doucet09}. 

The importance weight of every particle is given by:
\begin{align}
    \tilde{w}^{(i)}_t &= w^{(i)}_{t-1}\frac{{p(\vc{y}_t |\vc{x}^{(i)}_t)}{p(\vc{x}_t^{(i)}|\vc{x}^{(i)}_{t-1})}}{q(\vc{x}^{(i)}_t |\vc{x}^{(i)}_{t-1},\vc{y}_{t})}, \\
    w^{(i)}_{t} &= {\tilde{w}^{(i)}_t}/ {\sum\limits_{j=1}^N w^{(j)}_t}.
\end{align} \label{eq6}
%\tilde{w}^{(i)}_t =w^{(i)}_{t-1}\frac{{p(\vc{y}_t |\vc{x}^{(i)}_t)}{p(\vc{x}_t^{(i)}|\vc{x}^{(i)}_{t-1})}}{q(\vc{x}^{(i)}_t |\vc{x}^{(i)}_{t-1},\vc{y}_{t})},
%\end{equation}
It has been shown that PFs converge asymptotically, as $N\to\infty$, towards the optimal filter in the mean square error sense \cite{Doucet_2000}.

The selection of correct \emph{proposal distribution} $q(\vc{x}_t|\vc{x}_{t-1})$ is an important step in using PFs. Some studies have proposed to use either EKF \cite{Findeisen_1997_Thesis} \cite{Doucet_Report_1998} \cite{Doucet_2000} or UKF \cite{Merwe_U_PF_2004} to generate the importance PDF. At each step, an EKF or UKF is run for each particle to generate the mean and the covariance of the proposal PDF. Later, the particles are drawn from the newly found PDFs. The obvious advantage is that EKF and UKF take into account the most recent measurement while estimating the mean and the covariance. Further discussion on the proposal PDFs can be found in \cite{Doucet_2000}, \cite{Simandl_ProposalDensities_2007}, \cite{Merwe_SigmaPointKF_2004} and \cite{Merwe_Thesis_2004}.

Despite the selection of appropriate proposal densities, the sequential importance sampling algorithm may degenerate and never converge. The normalized weights of all but one particle degenerate to zero, referred to as \emph{sample impoverishment}. In order to avoid such a degeneracy problem, re-sampling is usually performed. Re-sampling will eliminate particles with low weight and multiply samples with high weights. Re-sampling algorithms are discussed in \cite{Bolic_ResamplingAlgo_2004}, \cite{Hol_Resampling_2006}.
The algorithm $1$ details the steps for particle filtering.
\begin{algorithm}
\caption{Particle filter}
\begin{algorithmic}
\STATE Generate $x_{0}^{(j)} \sim  q_0(x_0)$, then calculate $w_0^{i}=p(y_0|x_0^{i})$ and normalize the weights.
\FOR {$t=1,2,\cdots,T$ (where $T$ : time length)}
\FOR {$i=1,2,\cdots,N$ (where $N$ is the number of particles)}
\STATE Generate new samples from an accessible proposal distribution $x_{t}^{(i)} \sim  q_t(x_t)$.
\STATE Calculate the weights $w_t^{(i)}$ of $x_t^{(i)}$; then, normalize the weights.
\ENDFOR
\STATE Re-sample to obtain equally weighted particles $\{\vc{x}_t^{(i)}, \frac{1}{N}\}_{i=1}^N$.
\STATE Compute the weighted mean $\hat{x}_t= \sum\limits_{i=1}^{N}w_{t}^{(i)}x_{t}^{(i)} $.
\ENDFOR
\end{algorithmic}
\end{algorithm}
% % % %

\clearpage

\section{Constrained State Estimation} \label{section-constrained-CBE}
%\subsection{Constraint Definitions}
Many engineering applications, such as vision-based systems, chemical processes, target tracking, biomedical systems, navigation and robotics, can be modeled using state-space framework. In most of these dynamical system, the system state may be subject to constraints that arise from physical laws, natural phenomena, or model restrictions \cite{Simo06, Jun213, Jun113}. There may not be an easy and viable way to incorporate these constraints directly into the state-space model or the estimation framework \cite{AgSu04, YaBB06, HWHK12}.

We consider a set of constraints $C_t$, including linear and nonlinear, defined as:
\begin{align} \label{eq:Constraint_def}
    a_{t}\leq \phi_t(\hat{\vc{x}}_t)\leq b_t,
\end{align}
where $\phi_t$ represents the constraint function at time $t$. In the case when $\phi$ is an identify transformation, eq. (\ref{eq:Constraint_def}) reduces to a simple interval constraint on the mean of the state. The constraints $C_t$ can be hard or soft, where the estimation algorithms required to satisfy soft constraints approximately \cite{Simo10}.
%In this section, we summarize the available literature on state estimation when the unknown hidden state is required to satisfy certain constraints.

% We focus on the following constrained discrete state-space model:
% \begin{equation}\label{eq1-1}
% \left\{
%   \begin{array}{ll}
% x_{n+1}= f_n(x_n)+ w_n,\\
% y_n= h_n(x_n)+ v_n, ~~~\\

%   \end{array}
% \right.
% \end{equation}

\subsection{Linear Systems Subject to Constraints} \label{secLinearConstraints}
\subsubsection{Linear Constraints}
For linear Gaussian systems, linear constraints can be incorporated by directly within the KF framework as presented in eqs. \ref{eqnKFPrediction} and \ref{eqnKFUpdata}. The widely used methods are based on model reduction methodology \cite{Simon_CKF_2010}, \cite{Jiang_KF_2011}, \cite{Jiang_ROKF_2013}, pseudo-measurements, also called perfect measurement \cite{Gupta_Conf_2007}, \cite{Gupta_Doc_2009}, \cite{Teixeira_ControlsJournal_2009}, \cite{Duan_PseudoMeasurement_2013}, state estimate projection \cite{Gupta_Doc_2009}, \cite{Jiang_KF_2011}, \cite{Yang_Track_Fusion_2007}, and gain projection \cite{Gupta_Doc_2009}, \cite{Jiang_KF_2011}.

\subsubsection{Nonlinear Constraints}
For nonlinear constraints, a closed-form solution may not be possible, even for linear systems. The adopted methods rely on linear approximation of the nonlinear constraints using Taylor series expansion \cite{Yang_2ndOrder_2006}, \cite{Yang_Track_Fusion_2007} or on direct numerical optimization of the nonlinear problem \cite{Simon_CKF_2010}, \cite{Rao_MHE_2001}. Iterative application of the constraint linearization operation at each measurement time is used to get closer to the constraint satisfaction with each iteration \cite{Geeter_SCKF_1997}. Furthermore, PDF truncation methods have also been proposed that truncate Gaussian PDFs estimated by KF at the constraint bounds \cite{Simon_Book_2006}, \cite{Simon_PDFTrunc_2010}.

\subsection{Nonlinear Systems Subject to Constrains}
Linear constraints can be added directly in various estimation frameworks, i.e., EKF, UKF, or EnKF. The nonlinear constraints can linearized and then Incorporated in the filter.  

\subsubsection{Constrained State Estimation Using EKF}  
The system model (both dynamics and observation) and constraints can be linearized and used within the EKF framework for the constrained state estimation \cite{Ungarala07}. In some cases, the convergence properties of EKF can be improved by applying the filter iteratively for enforcing the constraint \cite{DeGeeter97, Prakash2014}. 

For the equality constraints, the measurement-argumentation approach can be used \cite{Alouani_PerfMeas_1993}, \cite{Chen_PM_EKF_1993}, \cite{Walker_PM_EKF_2006}, \cite{Ungarala_EKF_2007}. The measurement model can be augmented using the equality constraints and then EKF can be used for the linearized observation model. Obviously, measurement-argumentation approach is limited to equality constraints only.

%%%%
%%%%  Updated - March 11, 2022
%%%

Smoothing constrained kalman filter (SCKF) can be used with the EKF \cite{Geeter_SCKF_1997}. The SCKF constraints the system iteratively and the final constraint solution may not be accurate. This approach is only valid for equality constraints.

A modified extended Kalman filter has been introduced by Prakash \emph{et al.} to handle with the constraint imposed on the state estimation for a non-linear stochastic dynamic system \cite{Prakash2014}. Hence, Prakash \emph{et al.} proposed two schemes to modify the prior and posterior distributions based on generating samples from truncated multivariate normal distribution.

Recursive Nonlinear Dynamic Data Reconciliation (RNDDR) suitably combine advantages of both EKF and Nonlinear Dynamic Data Reconciliation (NDDR). The NDDR is a nonlinear optimization based strategy to estimate system parameters and states. Due to the optimization-based formulation of the problem, constraints on states and unknown parameters can be added in a natural way. However, the application of the NDDR for online estimation of states and parameters can be computationally prohibitive due to at each time-step, a nonlinear constrained optimization problem is solved.

Recently, Zixiao \emph{et al.} introduced a constrained dual extended Kalman filter algorithm that works in an alternating manner \cite{dualekf}. This algorithm is based on a parameter estimation and state prediction technique. The inequality constraints have been incorporated using optimization procedure. This proposed algorithm is more difficult to implement. However, it has the advantage of better convergence potential and algorithm stability.

In \cite{EKFG}, the inequality constraints have been incorporated into the EKF using a gradient projection method. The performance of the proposed technique has been tested using synthetic model based on Gaussian functions.

\subsubsection{Constrained State Estimation with UKF}
Many approaches have been proposed to incorporate the constraints within UKF.
During the update step, sigma points can be projected onto the constrained interval using sigma point projection approach \cite{Kandepu_C_UKF_2008}, \cite{Kandepu_C_UKF_J_2008}. This approach perform projection after generation of sigma points and then after passing sigma points through system dynamics.

Optimization in measurement-update has been proposed to reformulate the measurement-update step to integrate constraints within the standard UKF \cite{Kolas_C_UKF_2009}. This technique proposes to solve a quadratic or nonlinear optimization problem at each step of the algorithm.

Unscented Recursive Nonlinear Dynamic Data Reconciliation (URNDDR) is an extension of RNDDR where the EKF is replaced with the UKF \cite{Vachhani_RNDDR_2006}, \cite{Ungarala_URNDDR_Comments_2009}, \cite{Narasimhan_Reply_URNDDR_2009}. All types of constraints are taken into consideration by solving the constrained optimization problem for each sigma point. The algorithm may be computationally expensive because the optimization problem is solved for all sigma points. Recently Kadu, Mandela et al. introduced improvements to the URNDDR algorithm to address general constraints in the generation process of sigma points and computational issues \cite{Mandela_constrained_2012}.

Julier and LaViola proposed a two-step approach for tackling nonlinear equality constraints. Using the UKF approach, all the selected sigma points are projected onto the constrained surface individually in the first step, while in the second step the final estimate by the filter is again projected onto the constrained surface \cite{Julier_TwoStep_2007}. The authors presented a details discussion on the need for two-step projection for nonlinear constraints both for samples and their statistics (i.e., moments, which are expected value and covariance).

Teixeira \emph{et al.} extended the UKF for different types constraints. These approaches include equality constrained UKF, projected UKF, measurement-augmented UKF, constrained UKF, constrained-interval UKF, interval UKF, sigma point UKF, truncated UKF, truncated interval UKF, projected-interval UKF \cite{Teixeira_ControlsJournal_2009}, \cite{Teixeira_EqConst_Conf_2007}, \cite{Teixeira_Interval_UKF_Conf_2008}, \cite{Teixeira_JProccControl_2010}, \cite{Teixeira_Thesis_2008}. PDF truncation approach applied by Teixeira \emph{et al.} is applicable for linear interval constraints only \cite{Simon_Book_2006}.

Straka \emph{et al.} proposed a truncated unscented kalman filter (TUKF) algorithm to solve the non-linear and non gaussian system with constraints on the state estimation \cite{Straka_Truncation_2012}. The main idea of TUKF is to use the PDF truncation approach.

Recently, a truncated randomized unscented Kalman filter (TRUKF) was presented in \cite{staraka2013}. The TRUKF is based on the randomized unscented Kalman filter algorithm (RUKF) and a pdf truncation technique. The main idea of TRUKF scheme is to introduce a truncation step within RUKF  that was proposed for the unconstrained estimation problem. However, the application of TRUKF on synthetic data results in computational costs compared to UKF and TUKF (truncated UKF) \cite{staraka2013}.

Alireza \emph{et al.} proposed a novel approach called the constrained iterated unscented Kalman filter (CIUKF) \cite{alireza}. This approach combines the advantage of iterations and use of constraints to provide an accurate bounded dynamic state estimation. The constraints in the IUKF are incorporated by projection the sigma points that are outside the feasible region to the boundary of this region to obtain constrained sigma points.

Calabrese \emph{et al.} introduced an approach to integrate the constraints within the UKF based on two mainly approaches \cite{ACUKF}. In the first approach, all sigma points that violate the constrained region are moved onto the feasible region during the prediction step.
In the second approach, all transformed sigma points that violate the constrained region are projected to constraints boundary only when the updated state estimate exceeds the boundary in the correction step.

Variants of the algorithms seek to improve the performance and computational issues of the original UKF-based method under additional constraints \cite{Kandepu, Julier07, staraka2013}.

\subsubsection{Constrained state estimation using EnKF}

A constrained state estimation approach can also use the EnKF \cite{Prakash08a, dualkal}, where an initial ensemble of samples are drawn from a truncated normal distribution. For each iteration after the prediction step, a transformation is applied to project the violating samples onto the boundary.

Parakash \emph{et al.} presented constrained state estimation using the EnKF \cite{Prakash_C_EnKF_2010}, \cite{Prakash_C_EnKF_2008}. The authors propose to generate group of initial samples from a truncated normal distribution. Later, for each iteration after prediction step, a transformation is applied to project the violating samples on the boundary.

A Constrained Dual Ensemble Kalman Filter (dual C-EnKF) has emerged recently in \cite{dualkal}. The dual C-EnKF algorithm combines the C-EnKF algorithm proposed by Prakash in \cite{Prakash08a} for incorporating constraints within EnKF and the dual EnKF algorithm proposed in \cite{Moradkhani} to reduce the number of particles.

Raghu \emph{et al.} introduced two algorithms to incorporate the constraints into the EnKF \cite{enkf8}. The first algorithm uses the projection-based method. The second algorithm relies on the use of a technique for soft constrained covariance localization. Simulations results showed that the second proposed algorithm provide better estimation of the unknown states compared to the first proposed algorithm.

\subsubsection{Constrained state estimation using MHE}
It is evident from formulation of the MHE (\ref{eqnMHE}) that constraints can be incorporated into its framework in a natural way \cite{Rao_MHE_2001}, \cite{Rao_MHE_2002}, \cite{Rao_MHE_2003}, \cite{Haseltine_EKF_MHE_2005}, \cite{Rao_PhDThesis_2000}. However, there are multiple issues with MHE framework, i.e., 1) the computational effort especially for nonlinear optimization problem (constraints and/or objective functions) \cite{Rao_MHE_2002}, \cite{Rao_PhDThesis_2000}, \cite{Rawlings_PF_MHE_2006}, \cite{Zavala_Fast_MHE_2008}; 2) calculation of the arrival cost \cite{Ungarala_DirSmp_PF_2011}; 3) Gaussian assumption for densities that results in simplification of the relations, i.e., from Eq. (\ref{eqnMHE1}) to (\ref{eqnMHE}); and 4) selection of optimal horizon ($ h $) size to balance performance and computational load \cite{Ungarala_DirSmp_PF_2011}.

Recently, Garcia Tirado \emph{et al.} introduced an approach for constrained estimation problem depends on the MHE and the game theoretical approach to the $\mathcal{H}_\infty$ filtering with constraint handling \cite{MHEC}.
The theoretical of the proposed approach with constraints mainly based on a modified Lyapunov theory for optimization-based systems.

%\cite{MHEC} important to add it the name of paper beys2017

It is important to emphasize that almost all methods that we present for the constrained state estimation have an underlying assumption of linearity or Gaussianity, an unrealistic presumption in most real-world applications. Moreover, the presence of constraints, such as bounds, on the states implies that the conditional state densities are non-Gaussian. Furthermore, the UKF and EnKF-based methodologies systematically constrain all their sigma points and ensemble samples with no mathematical ground or justification.

\section{Constrained State Estimation Using Particle Filters}  \label{section-constrained-PF}
Particle filters are widely used for latent state estimation/tracking in dynamic systems where systems dynamics or observation models are nonlinear, or the system/observation noise are non-additive or follow non-Gaussian distributions \cite{Doucet09}. The technique of PFs is based on powerful sampling that is aimed to find an optimal estimate by exploiting a set of random weighted samples called the particles. These particles are used to approximate the posterior density of the state and later find the statistics of interest \cite{Doucet09}. Due to the complex nature of computations in PFs, it is not straightforward to incorporate constrains on the latent state. Systematic efforts to incorporate constraints imposed on the unknown state in PFs are limited and heuristic in nature.
\subsection{Acceptance/Rejection Approach}
An acceptance-rejection approach was proposed for nonlinear inequality constraints \cite{Lang_Accept_Reject_2007, standoff}. This approach focused on retaining particles that fell within the constrained interval and rejecting all violating constraint region. However, their approach does not make any assumption on the distributions and can guarantee the validity of particles, and thus retains the general properties of the particle filter. Besides, in certain cases, the number of particles may reduce which may further lead to a decrease in estimation accuracy and computationally efficiency.

In such cases, most of the particle may violate the constraint and the algorithm may fail. Also, unconstrained sampling from followed by verification against constraints (especially nonlinear) may be computationally more demanding than sampling directly from the constrained region  \cite{Ungarala_DirSmp_PF_2011, Lang_Accept_Reject_2007, standoff}.

%PoDeT has been used until now to solve may engineering problems such as: the standoff target tracking guidance using constrained particle filter for multiple UAVs. The constraints has been incorporated with the acceptance–rejection process \cite{standoff}.
\subsection{Optimization/Projection Based Approach}
Shao \emph{et al.} presented a two stages approach to deal with constraints \cite{ XShao10}. In the first stage, a set of particles were drawn randomly without considering state constraints. In the second stage, all particles which did not satisfy constraints, were projected into the feasible region using an optimization formulation. However, by applying optimization method to impose the particle to be within the constraining interval, the obtained particles are no longer considered as representative samples of the posterior distribution of the state. At every sampling instant, solving multiple optimization problems may make the algorithm computationally expensive.

Xiong \emph{et al.} proposed an adaptive constrained particle filter (ACPF) algorithm that uses all particles to accurately estimate the state \cite{Xiong2014}. The approach is based on sample size testing that allows calculation of the number of particles needed to obtain a desired state estimate. Xiong \emph{et al.} combined the sample size test with the generic particle filter to deal with constraints and to address the particle number problem in PF. The simulation results showed that, in terms of root mean square error (RMSE) performance, convergence, and running time, the ACPF approach exhibits a better estimation compared to the constrained PF proposed in \cite{ XShao10} and unconstrained PF.

Li \emph{et al.} used a series of constrained optimization to incorporate the inequality constraints into the auxiliary particle filter by modifying the priori pdf \cite{Li}. The auxiliary particle filter consists of resampling and sampling steps at each time step. In the resampling steps, it selected particles with a lower likelihood and/or far from the feasible region. It then performed a series of constrained optimization to transform the center of transition distribution into a feasible region.

Recently, Hongwei \emph{et al.} introduced a constrained multiple model particle filtering (CMMPF) method to solve the constrained high dimensional state estimation problems \cite{CPF8}. The proposed approach divided the problem of target tracking into two sub-problems: i) motion model estimation and model-conditioned state filtering as stated in the Rao–Blackwellised theorem; ii) The hidden state estimation is formulated using the multiple switching dynamic models in a jump Markov system framework. In order to incorporate the constraints within the proposed approach, a modified sequential importance resampling (MSIR) method based on a series of optimizations is used to generate model particles that can be constrained in the feasible region.

\subsection{ Constrained Importance Distributions}
The particles can be drawn from a proposal distribution having support on the constrained region only. There are different approaches to incorporating constraints in the PF using the proposal distribution presented in literature. Density truncation can be performed analytically in case of multivariate Gaussian distribution \cite{Simon_PDFTrunc_2010}, \cite{Gates_GHKSimulator_2006}. Perfect Monte Carlo simulations can be used to estimate first two moments of the truncated proposal distribution. Samples are drawn from a distribution of interest and all constraint-violating samples are rejected. Leftover samples are used to estimate mean and covariance of the truncated PDF.

The importance distribution can be constrained also, i.e., in the sampling step, particles are drawn from an importance distribution which has its support on the constrained region only \cite{Prakash08b, Prakash_Choice_IF_2011}. Constrained-EKF, constrained-UKF and or constrained-EnKF are used at each iteration to generate constrained proposal distributions. Specifically, a constrained-filter (EKF, UKF or EnKF) is used for each particle and an estimate of mean and covariance of constrained distribution is found. Particles are drawn from the newly found proposal distribution. An analytical solution for PDF truncation is also proposed \cite{Prakash08b, Prakash_Choice_IF_2011}.

Straka \emph{et al.} \cite{Straka_Truncation_2012} proposed a truncated unscented particle filter to incorporate the constraints in the unscented particle filter. Proposal density is generated using the UKF and before sampling particles from it, a truncation procedure, in accordance with constraints on the system, is performed to form a truncated proposal density. The truncated proposal density is formed using perfect Monte Carlo or importance-sampling. The truncation procedure produces mean and variance of the proposal density, which is assumed to be Gaussian.

On the other hand, Yu \emph{et al.} proposed a truncated unscented particle filter to handle nonlinear constraints \cite{yu2014}. Their technique combines both PF based accept/reject approach and the unscented Kalman filter. Authors start by drawing the importance distribution for sampling new particles using the unscented Kalman filter and then applied the truncated probability estimation using accept/reject approach. However, this technique also used the accept/reject approach and thus suffers from same limitations as mentioned earlier.

Similarly, Pirard \emph{et al.} proposed  two approaches to solve the problem of target tracking in the presence of constraints \cite{Pirard2013}. The first approach was based on the extension of the Rao-Blackwellized PF (RBPF) to handle hard constraints. The technique of RBPF or marginalized PF can be used only when the state can be divided into two parts such as a linear part and a non-linear part, and where the constraints only depend on the non-linear part. This approach used the accept/rejection technique that may lead to a reduction the estimation accuracy. The second approach was built on the proposal distributions adapted to the constraints. This approach drew a proposal distribution which guaranteed that particles were drawn from constrained regions. This could be quite ineffective when a substantial part of the drawn particles is located in outside constrained regions.

Ungarala \emph{et al.} proposed constrained Bayesian state estimation using a cell filter \cite{Ungarala08}, where a Markov chain is constructed by sampling the dynamics over constraints. However, this approach is limited to low dimensional systems due to exponentially increasing memory requirements of the state transition operator with the state dimension. In addition, Ungarala \cite{Unga11} proposed a direct sampling particle filter for linear and nonlinear constraints, however, its applicability is limited to Gaussian assumption for all pdfs.

Zhao et \emph{et al.} \cite{Zhao_3Stage_2012} proposed three strategies for constrained state estimation using particle filters. The first strategy ensures that the samples are drawn from the constrained region using a constrained inverse transform technique. Using bounds (interval constraint) on the state vector, bounds on the process noise are found. Process noise samples are drawn from the constrained commutative distribution function (CDF). In the second strategy, an acceptance/rejection scheme is used after re-sampling and all violating particles are rejected. The third strategy deals with particles after resampling.
Violating particles are deleted and non-violating particles are regenerated to ensure that there are more non violating particles in the final estimate.
There are a few problems with this approach. First, it may not be possible to find the CDF of the noise in the first strategy. Also, the constraints on the state may not be of the interval-type, i.e, may have a nonlinear form, then the first strategy is not applicable. For the second strategy, the acceptance/rejection scheme may result in rejection of most of the particles in the worst case, leading to failure of the particle filter. The third strategy generates multiple number of state estimates (one for each deletion/regeneration) and an optimization problem is solved to find the right state estimate out of all available. Essentially, another optimization problem has to be solved which may be computationally expensive.

In addition, Zhao \emph{et al.} proposed three strategies for constrained state estimation \cite{Zhao2014}. The first strategy focused on constrained prior particles using inverse and Gibbs sampling. This strategy used first the Gibbs sampling to compute the constraint interval of each variable in the state vector, and then generated valid particles using constrained inverse transform sampling. However, the generation of each particle required computationally expensive processes of constrained region calculation and then and then generation of particles. In the second proposed strategy, an accept/reject scheme is used to constraint the posterior particles. This scheme may result in the rejection of many particles in the worst case, leading to failure of the PF. The third proposed strategy imposed constraints on the state by scaling the weights of the particles. However, the scaling of weights in this way contradicts basic PF theory.

In summary constrained state estimation includes accept/reject approaches, projection/optimization based methods, and constrained importance sampling distribution. We referred to these schemes as “POintwise DEnsity Truncation” (PoDeT) \cite{Amor2016}. All PoDet methods impose constraints on all particles of the PF and thus constrain the posterior distribution of the state estimate rather than its mean. This may lead to more stringent conditions than actually desired and may also result into possibly irrelevant conditions than the original constraints \cite{Amor2017}.

\subsection{Challenges in Constrained Particle Filtering}
The particle filtering represents the state-of-the-art for estimation in nonlinear/non-Gaussian dynamical systems; however, incorporation of constraints on the hidden state (e.g., non-negativity) is a challenging problem.
%Generally, the state estimates by the PFs are constrained by simply constraining all particles leading to constraining the support of the posterior density.

\subsubsection{On The Convergence of Constrained Particle Filtering}
By constraining every particle in the PF, the PoDeT approach will always result in an estimation error of the posterior density unless the density has a bounded support. In particular, if the unconstrained distribution naturally satisfies the constraints and has unbounded support, the PoDeT will fail in rendering the unconstrained density. An evaluation of the PoDeT estimation error is discussed in \cite{Amor2017}, where we derived the optimality bounds of the PoDeT approach.
We have shown that the estimation error is bounded by the area of the state posterior density that does not include the constraining interval. Specifically, if the density is well localized, i.e., most of the unconstrained posterior density were inside the constraining interval, then the PoDeT error will be upper bounded. On another hand, if the density is not well-localization, i.e., the constrained interval was did not contained most of the posterior distribution, then the PoDeT error will be bounded from below. In particular, if the constraining interval covers a small area of the density, then the PoDeT error will be large \cite{Amor2017}.

%We have demonstrated that the PoDeT approach results in a bounded estimation error to the posterior density of the state if the target density is \emph{well-localized} in the interval $[a, b]$, i.e., when the unconstrained posterior density is already inside the constraining interval \cite{Amor2017}. A density is \emph{well-localized} within the interval $[a, b]$ if the probability of this interval is close to unity. Mathematically, the characterization of the ``localization" of a pdf with respect to an interval can be given in terms of the probability of the interval: if $Pr\{[a, b]\} \geq 1 -\eta$, where $0 \leq \eta \ll 1$ is a small number, the density is considered well-localized \cite{Amor2017}. In particular, the area under the pdf that is bounded by the interval $[a, b]$, represents an important parameter for controlling the estimation error of PoDeT. Intuitively, if high probability areas of the density are within the restriction interval, then the conditional mean estimate will be close to the truncated density at the support. In this concept, the estimated error in the posterior density is small and can be determined using the region under the tails of the well-localized density, i.e., the pdf region in the interval  {$]-\infty, a[ \cup ]b,\infty[$}.
\subsubsection{Mean Density Truncation}
Based on the PoDeT estimation error results, the \emph{mean density truncation} (MDT) method has been recently proposed in \cite{Ebinger15}. MDT deals with errors of the PoDeT by proposing a new strategy to satisfy the constraint on the conditional mean estimate rather than the posterior density itself \cite{Ebinger15}. The authors generated $N$ unconstrained particles from the importance distribution as in the standard PF. If these $N$ weighted particles satisfied the constraints on the mean, the constrained state estimate is considered optimal. If not, an $(N +1)$st particle was drawn from a high probability region to enforce constraint on the weighted mean. In case if $(N +1)$st particle was not sufficient to guarantee the constraints on the conditional mean, authors generated additional particles iteratively one by one ($2$, then $3$, then $4$ ..., etc.) until the constraints were satisfied. One drawback of this approach was evident when the proposal distribution was poorly chosen. In that case, it may take a large number of particles to move the mean of the conditional distribution into the feasible region. Consequently, the iterative technique of generating particles one by one may make the algorithm inefficient and computationally expensive.

Later, authors introduced systematic and inductive procedure to ensure that the constraints were satisfied with the generation of $N$ particles, referred to as the \emph{Iterative Mean Density Truncation} (IMeDeT) \cite{Amor2016}. It is important to notice that, this perturbation approach remains very different from the PoDeT approach where the original constraint is imposed on every particle, while the $N$-particle IMeDeT enforces the desirable constraint only on the conditional mean estimate. However, the only drawback of IMeDeT is that inductively choosing particles $j=1, \ldots, N$ to satisfy constraints on the conditional mean at every time step may lead to computationally expensive.

Recently, we addressed the limitations of MDT and IMeDeT by proposing a new strategy based on perturbing the unconstrained density with only one particle \cite{Amor2018}. We referred to this technique as \emph{Mean Density Truncation}(MeDeT). In MeDeT, we choose one particle in a special way to satisfy the constraints on the mean and construct a sequence of densities satisfying constraints \cite{Amor2018}. We start by generating $N$ unconstrained particles from the importance sampling distribution. If the conditional mean estimate using the N-order approximation satisfies the constraints, we retain these particles. Otherwise, after the resampling step, we remove a particle with the lowest weight and located closest to the boundary of the feasible region and replace it with another particle that is drawn from the high probability region. The process of remove/add one particle can be viewed as the "minimal perturbation" of the unconstrained density with using only one particle.

Keeping in view the way constrained are implemented during filtering, we can classify constraints as \emph{soft constraints} and \emph{hard constraints}. In the techniques of MDT, IMeDeT and MeDeT, the state satisfies the constraint on the conditional mean, so the constraints are considered soft constraints. On the other hand, for PoDeT, the constraints are enforced on each particle and are considered hard constraints.

We hope that this paper encourages further research in the development of more algorithms that constrain the state estimate rather than the density itself.

\section{Conclusion}\label{section-conclusion}
This paper reviews the advances approaches to incorporate the constraints within state estimation.
We provided a critical review of constrained Bayesian state estimation algorithms using Kalman filter for linear systems and extended Kalman filter, unscented Kalman filter, ensemble Kalman filter, moving horizon estimation and particles filters for nonlinear systems.

\section*{Acknowledgment}
 This work was supported by the National Science Foundation under Grants NSF CCF-1527822 and NSF ACI-1429467.

\bibliographystyle{ieeetran}
\bibliography{CBayEst}
\end{document}